\documentclass[preprint,12pt]{elsarticle}
\usepackage{graphicx}
\usepackage{amsmath}
\usepackage{amssymb}
\usepackage{bm}% bold math
\begin{document}

%\begin{frontmatter}

\title{Suppression of oscillations by L\'evy noise}

\author[iaf,sumdu]{A. I. Olemskoi\corref{cor2}}
\ead{alex@ufn.ru}
\address[iaf]{Institute of Applied Physics, Nat. Acad. Sci. of Ukraine,
\break 58 Petropavlovskaya St., 40030 Sumy, Ukraine}
\address[sumdu]{Sumy State University, 2 Rimskii-Korsakov St., 40007 Sumy, Ukraine}
\author[sumdu]{S. S. Borysov\corref{cor1}}
\ead{stasmix@gmail.com}
\author[sumdu]{I. A. Shuda}
\ead{shudaira@mail.ru}

\cortext[cor1]{Corresponding author}
\cortext[cor2]{Principal corresponding author}

\date{}

\begin{abstract}
We find analytical solution of pair of stochastic equations with arbitrary
forces and multiplicative L\'evy noises in a steady-state nonequilibrium case.
This solution shows that L\'evy flights suppress always a quasi-periodical
motion related to the limit cycle. We prove that such suppression is caused by
that the L\'evy variation $\Delta L\sim(\Delta t)^{1/\alpha}$ with the exponent
$\alpha<2$ is always negligible in comparison with the Gaussian variation
$\Delta W\sim(\Delta t)^{1/2}$ in the $\Delta t\to 0$ limit. Moreover, this
difference is shown to remove the problem of the calculus choice because
related addition to the physical force is of order $(\Delta
t)^{2/\alpha}\ll\Delta t$.
\end{abstract}

\begin{keyword}
L\'evy noise; Stationary state; Limit cycle \PACS {02.50.Ey,
05.40.Fb, 82.40.Bj}
\end{keyword}
%\end{frontmatter}
\maketitle

\section{Introduction}\label{Sec.1}

It is known crucial changing in behavior of the systems that display
noise-induced \cite{2,2a} and recurrence \cite{3a,3} phase transitions,
stochastic resonance \cite{4a,4}, noise induced pattern formation \cite{a,b},
noise induced transport \cite{c,2a} etc. is caused by interplay between noise
and non-linearity (see Ref. \cite{SSG}, for review). Noises of different origin
can play a constructive role in dynamical behavior such as hopping between
multiple stable attractors \cite{e,f} and stabilization of the Lorenz attractor
near the threshold of its formation \cite{d,g}. This type of behavior is
inherent in finite systems where examples of substantial alteration under
effect of intrinsic noises give epidemics \cite{11}--\cite{13}, predator-prey
population dynamics \cite{5,6}, opinion dynamics \cite{10}, biochemical clocks
\cite{15,16}, genetic networks \cite{14}, cyclic trapping reactions \cite{9} et
cetera.

Above pointed out phase transitions present the simplest case, when joint
effect of both noise and non-linearity arrives at non-trivial fixed point
appearance only on the phase-plane of the system states. In this consideration,
we are interested in studying much more complicated situation, when stochastic
system may display oscillatory behavior related to the limit cycle appearing as
a result of the Hopf bifurcation \cite{17,18}. It has long been conjectured
\cite{19} that in some situations the influence of noise would be sufficient to
produce cyclic behavior \cite{20}. Moreover, it has been shown that excitable
\cite{21a}, bistable \cite{21b} and close to bifurcations \cite{21c} systems
display oscillation behavior, whose adjacency to ideally periodic signal
depends resonantly on the noise intensity \cite{21d} (due to this reason, such
oscillations were been called coherence resonance \cite{21a} or stochastic
coherence \cite{SSG}).

Characteristic peculiarity of above considerations is that all of them are
restricted by studying the Gaussian noise effect, while such a noise is a
special case of the L\'evy stable process (the principle difference of these
noises is known \cite{rev} to consist in the form of the probability
distribution that exhibits the asymptotic power-law decay in the latter case
and decays exponentially in the former one). Nowadays, anomalous diffusion
processes associated with the L\'evy stable noise are attracting much attention
in a vast variety of fields not only of natural sciences (physics, biology,
earth science, and so on), but of social sciences such as risk management,
finance, etc.

In the context of physics, recent investigation \cite{0} has shown that joint
effect of both non-linearity and L\'evy noise may cause the occurrence of
genuine phase transitions which relates to a fixed point on the phase-plane of
the system states. In this connection, natural question arises: may be
displayed a self-organized quasi-periodical behavior related to the limit cycle
by a system driven by the L\'evy stable noise? This work is devoted to the
answer to above question within analytical study of two-dimensional stochastic
system.

The paper is organized as follows. In Section \ref{Sec.2}, we consider pair of
stochastic equations with arbitrary forces and multiplicative L\'evy noises to
obtain their analytical solution in a steady-state nonequilibrium case. This
allows us to conclude in Section \ref{Sec.3} that opposite to the Gaussian
noises the L\'evy flights suppress always a quasi-periodical motion related to
the limit cycle. Since equation, governing behavior of stochastic system driven
by multiplicative L\'evy stable noise, are very complicated \cite{Yanovsky} and
moreover their derivation is now in progress \cite{Denisov}, we complete our
consideration with Appendix A containing details of derivation of the
Fokker-Planck equation. Moreover, to demonstrate that a closed consideration of
the L\'evy processes is achieved only within the Fourier representation we set
forth a scheme related to the appropriate stochastic space in Appendix B.

\section{Statistical picture of limit cycle}\label{Sec.2}

According to the theorem of central manifold \cite{17}, to achieve a closed
description of a limit cycle it is enough to use only two degrees of freedom
related to some stochastic variables $X_i$, $i=1,2$. In this way, stochastic
evolution of the system under investigation is defined by the Langevin
equations \cite{23}
\begin{equation}
{\rm d}X_i=f_i{\rm d}t+g_i{\rm d}L_i,\quad i=1,2
 \label{1}
\end{equation}
with arbitrary forces $f_i=f_i(x_1,x_2)$ and noise amplitudes
$g_i=g_i(x_1,x_2)$ being functions of both variables $x_i$, $i=1,2$; stochastic
terms are related to the L\'evy stable processes $L_i=L_i(t)$. Within the \^Ito
calculus, these processes are determined by the elementary characteristic
function
\begin{equation}
\left<{\rm e}^{{\rm i}k_i{\rm d}X_i}\right>:={\rm e}^{\mathcal{L}_i{\rm d}t}
 \label{2}
\end{equation}
with increments $\mathcal{L}_i=\mathcal{L}_i(k_1,k_2;x_1,x_2)$ whose expression
\cite{Yanovsky}
\begin{equation}
\mathcal{L}_i={\rm i}k_i\left(f_i+\gamma_i
g_i\right)-|m_ig_ik_i|^{\frac{\alpha}{2}}{\rm e}^{-{\rm
i}\varphi_i(\frac{\alpha}{2})}\sum\limits_{j=1}^2\left|m_jg_jk_j\right|^{\frac{\alpha}{2}}{\rm
e}^{-{\rm i}\varphi_j(\frac{\alpha}{2})}
 \label{3}
\end{equation}
follows from Eq.(\ref{25a}). Hereafter, we use asymmetry angles $\varphi_i$ and
moduli $m_i$ defined by the equalities
\begin{equation} \label{12}
\begin{split}
\tan\left[\varphi_i(\alpha)\right]&=\beta_i{\rm
sgn}(g_ik_i)\tan(\pi\alpha/2),\\ m_i^\alpha&=
\sqrt{1+\beta_i^2\tan^2(\pi\alpha/2)};
\end{split}
\end{equation}
everywhere, the L\'evy index $\alpha\in(0,2)$ characterizes the asymptotic tail
$x_i^{-(\alpha+1)}$ of the L\'evy stable distribution at $1\ne\alpha<2$ (the
case $\alpha=2$ relates to the Gaussian distribution), parameters
$\beta_i\in[-1,+1]$ define the distribution asymmetry, location parameters
$-\infty<\gamma_i<+\infty$ denote the mean values of stochastic variables $X_i$
at $\alpha>1$, and the angular brackets denote averaging over L\'evy noises.

As is shown in Appendix A, the Fourier transformed probability distribution
function
\begin{equation}
\widetilde{P}(k_1,k_2;t)\equiv\mathcal{F}\{P(x_1,x_2)\}(k_1,k_2;t)
:=\iint\limits_{-\infty}^{+\infty}{\rm d}x_1{\rm d}x_2~P(x_1,x_2;t) {\rm
e}^{{\rm i}\left(k_1x_1+k_2x_2\right)}
 \label{5}
\end{equation}
is governed by the Fokker-Planck equation
\begin{equation} \label{6}
\frac{\partial\widetilde{P}}{\partial t}=\sum\limits_{i=1}^{2}\left[{\rm
i}\left(f_i+\gamma_ig_i\right)k_i-|m_ig_ik_i|^{\frac{\alpha}{2}}{\rm e}^{-{\rm
i}\varphi_i(\frac{\alpha}{2})}\sum\limits_{j=1}^2\left|m_jg_jk_j\right|^{\frac{\alpha}{2}}{\rm
e}^{-{\rm i}\varphi_j(\frac{\alpha}{2})}\right]\widetilde{P}.
\end{equation}
Characteristically, being Fourier transformed, r.h.s. of this equation depends
on the wave vector components $k_1$ and $k_2$, while both forces
$f_i=f_i(x_1,x_2)$ and multiplicative noise amplitudes $g_i=g_i(x_1,x_2)$ are
dependent on the coordinate components $x_1$ and $x_2$.

According to the continuity equation (\ref{23a}), components of the
steady-state probability flux are obeyed to the condition $\sum_i\partial
J_i/\partial x_i=0$ which means the first component $J_1=J_1(x_2)$ is a
function of the only variable $x_2$, and vice-versa for the second component
$J_2=J_2(x_1)$. Then, within the Fourier representation, the system behaviour
is defined by the equations
\begin{equation} \label{7}
\begin{split}
&\left\{\left(f_1+g_1\gamma_1\right)+{\rm i}|m_1g_1|^{\frac{\alpha}{2}}{\rm
e}^{-{\rm
i}\varphi_1\left(\frac{\alpha}{2}\right)}|k_1|^{\frac{\alpha}{2}-2}k_1\right.\\
&\left.\times \left[\left|m_1g_1k_1\right|^{\frac{\alpha}{2}}{\rm e}^{-{\rm
i}\varphi_1\left(\frac{\alpha}{2}\right)}+
\left|m_2g_2k_2\right|^{\frac{\alpha}{2}}{\rm e}^{-{\rm
i}\varphi_2\left(\frac{\alpha}{2}\right)}\right]\right\}\widetilde{P}=2\pi
J_1(k_2)\delta(k_1),
\end{split}
\end{equation}
\begin{equation} \label{8}
\begin{split}
&\left\{\left(f_2+g_2\gamma_2\right)+{\rm i}|m_2g_2|^{\frac{\alpha}{2}}{\rm
e}^{-{\rm
i}\varphi_2\left(\frac{\alpha}{2}\right)}|k_2|^{\frac{\alpha}{2}-2}k_2\right.\\
&\left.\times\left[\left|m_1g_1k_1\right|^{\frac{\alpha}{2}}{\rm e}^{-{\rm
i}\varphi_1\left(\frac{\alpha}{2}\right)}+
\left|m_2g_2k_2\right|^{\frac{\alpha}{2}}{\rm e}^{-{\rm
i}\varphi_2\left(\frac{\alpha}{2}\right)}\right]\right\}\widetilde{P}=2\pi
J_2(k_1)\delta(k_2).
\end{split}
\end{equation}
Since the pair of these equations determines a single distribution function
$\widetilde{P}(k_1,k_2)$, the consistency condition
\begin{equation} \label{condition}
\begin{split}
\left[\left(f_1+g_1\gamma_1\right)+{\rm i}{\rm e}^{-{\rm
i}\varphi_1(\alpha)}|m_1g_1|^\alpha|k_1|^{\alpha-2}k_1\right]
\delta(k_2)J_2(k_1)\\ =\left[\left(f_2+g_2\gamma_2\right)+{\rm i}{\rm e}^{-{\rm
i}\varphi_2(\alpha)}|m_2g_2|^\alpha|k_2|^{\alpha-2}k_2\right]
\delta(k_1)J_1(k_2)
\end{split}
\end{equation}
should be kept to restrict the choice of the probability flux components
$J_1(k_2)$ and $J_2(k_1)$.

Multiplying Eq.(\ref{7}) by the factor $|m_2g_2|^{\frac{\alpha}{2}}{\rm
e}^{-{\rm i}\varphi_2\left(\frac{\alpha}{2}\right)}$ and Eq.(\ref{8}) by
$|m_1g_1|^{\frac{\alpha}{2}}{\rm e}^{-{\rm
i}\varphi_1\left(\frac{\alpha}{2}\right)}$ and then subtracting results, one
obtains
\begin{equation} \label{9}
\begin{split}
\left\{F+{\rm i}|m_1m_2g_1g_2|^{\frac{\alpha}{2}}{\rm e}^{-{\rm
i}\left[\varphi_1\left(\frac{\alpha}{2}\right)+\varphi_2\left(\frac{\alpha}{2}\right)\right]}\right.\\
\left.\times \left[\left|m_1g_1k_1\right|^{\frac{\alpha}{2}}{\rm e}^{-{\rm
i}\varphi_1\left(\frac{\alpha}{2}\right)}+
\left|m_2g_2k_2\right|^{\frac{\alpha}{2}}{\rm e}^{-{\rm
i}\varphi_2\left(\frac{\alpha}{2}\right)}\right]\left(|k_1|^{\frac{\alpha}{2}-2}k_1
-|k_2|^{\frac{\alpha}{2}-2}k_2\right)\right\} \widetilde{P}\\ =2\pi
\left[J_1(k_2)\delta(k_1)|m_2g_2|^{\frac{\alpha}{2}}{\rm e}^{-{\rm
i}\varphi_2\left(\frac{\alpha}{2}\right)}-J_2(k_1)\delta(k_2)|m_1g_1|^{\frac{\alpha}{2}}{\rm
e}^{-{\rm i}\varphi_1\left(\frac{\alpha}{2}\right)}\right]
\end{split}
\end{equation}
where one denotes
\begin{equation} \label{9aa}
F\equiv\left(f_1+\gamma_1g_1\right)|m_2g_2|^{\frac{\alpha}{2}}{\rm e}^{-{\rm
i}\varphi_2\left(\frac{\alpha}{2}\right)}-\left(f_2
+\gamma_2g_2\right)|m_1g_1|^{\frac{\alpha}{2}}{\rm e}^{-{\rm
i}\varphi_1\left(\frac{\alpha}{2}\right)}.
\end{equation}
The equation (\ref{9}) yields the explicit form of the probability distribution
function
\begin{equation} \label{10}
\begin{split}
P\left(x_1,x_2\right)&=\int\limits_{-\infty}^{+\infty}\frac{{\rm d}k_2}{2\pi}
\frac{J_1(k_2)|m_2g_2|^{\frac{\alpha}{2}}{\rm e}^{-{\rm
i}\left[k_2x_2+\varphi_2\left(\frac{\alpha}{2}\right)\right]}}{F_2-{\rm
i}|g_1|^{\frac{\alpha}{2}}|m_2g_2|^{\alpha}{\rm e}^{-{\rm
i}\varphi_2\left(\alpha\right)}|k_2|^{\alpha-2}k_2}\\
&-\int\limits_{-\infty}^{+\infty}\frac{{\rm
d}k_1}{2\pi}\frac{J_2(k_1)|m_1g_1|^{\frac{\alpha}{2}}{\rm e}^{-{\rm
i}\left[k_1x_1+\varphi_1\left(\frac{\alpha}{2}\right)\right]}}{F_1+{\rm
i}|g_2|^{\frac{\alpha}{2}}|m_1g_1|^{\alpha}{\rm e}^{-{\rm
i}\varphi_1\left(\alpha\right)} |k_1|^{\alpha-2}k_1}
\end{split}
\end{equation}
where effective forces $F_{1,2}$ are determined by Eq.(\ref{9aa}) at
$m_{2,1}=1$ and $\varphi_{2,1}=0$.

In the case of constant values of the probability flux within the state space
$x_1,x_2$, the Fourier transforms related are $J_1(k_2)=2\pi
J_1^{(0)}\delta(k_2)$ and $J_2(k_1)=2\pi J_2^{(0)}\delta(k_1)$ with
$J_i^{(0)}={\rm const}$. Then, the consistency condition (\ref{condition})
takes the form
$\left(f_1+g_1\gamma_1\right)J_2^{(0)}=\left(f_2+g_2\gamma_2\right)J_1^{(0)}$,
the effective force (\ref{9aa}) is
$F_0=\left(f_1+\gamma_1g_1\right)|g_2|^{\frac{\alpha}{2}}-\left(f_2
+\gamma_2g_2\right)|g_1|^{\frac{\alpha}{2}}$, and the probability density
(\ref{10}) reads
\begin{equation} \label{11cc}
P=\frac{J_1^{(0)}|g_2|^{\frac{\alpha}{2}}-J_2^{(0)}|g_1|^{\frac{\alpha}{2}}}{F_0}.
\end{equation}
To create a limit cycle this distribution function should diverges on a closed
curve, so that the effective force equals $F_0=0$. Together with the
consistency condition, this equation gives
\begin{equation} \label{11c}
\frac{J_1^{(0)}}{J_2^{(0)}}=\frac{f_1+\gamma_1g_1}{f_2
+\gamma_2g_2}=\left|\frac{g_1}{g_2}\right|^{\frac{\alpha}{2}}.
\end{equation}
But these equalities mean that the numerator of the probability density
(\ref{11cc}) disappears also. As a result, we conclude the limit cycle creation
is impossible for a stationary non-equilibrium state with both probability flux
components $J_1(x_1,x_2)$ and $J_2(x_1,x_2)$ being constant.

To calculate integrals in Eq.(\ref{10}) for arbitrary dependencies $J_1(k_2)$
and $J_2(k_1)$ it is convenient to write $|k|={\rm sgn}(k)k={\rm e}^{{\rm
i}\pi\theta(-k)}k$ where $\theta(k)$ denotes the Heaviside step function. Then,
one has $|k|^{\alpha-2}k={\rm e}^{-{\rm
i}\pi\theta(-k)(2-\alpha)}k^{\alpha-1}$, and the pole points of integrands in
Eq.(\ref{10}) are expressed with the equality
\begin{equation} \label{11}
\begin{split}
K_{1,2}&=\left(\frac{F_{1,2}}{|m_{1,2}g_{1,2}|^\alpha|g_{2,1}|^{\frac{\alpha}{2}}}
\right)^{\frac{1}{\alpha-1}} \\ &\times\exp\left\{{\rm i
}\frac{\varphi_{1,2}(\alpha)+(2-\alpha)\pi\theta(-\Re K_{1,2})+(\pi/2){\rm
sgn}(\Im K_{1,2})}{\alpha-1}\right\}.
\end{split}
\end{equation}
Due to sign-changing term $(\pi/2){\rm sgn}(\Im K_{1,2})$ in the exponent the
$K_{1,2}$ poles are located on opposite half-planes of complex variables
$k_{1,2}$. Making use of the power series expansion
\begin{equation} \label{15}
k^{\alpha-1}=K^{\alpha-1}\left(1+\frac{k-K}{K}\right)^{\alpha-1}\approx
K^{\alpha-1}+(\alpha-1)K^{\alpha-2}(k-K)
\end{equation}
allows us to reduce the integrands in Eq.(\ref{10}) to a pole form. However, we
can not close the integration contours around both upper and lower complex
half-planes of the $k$ variable since integrands related contain absolute
magnitudes.

\begin{figure}[!htb]
\centering
\includegraphics[width=120mm]{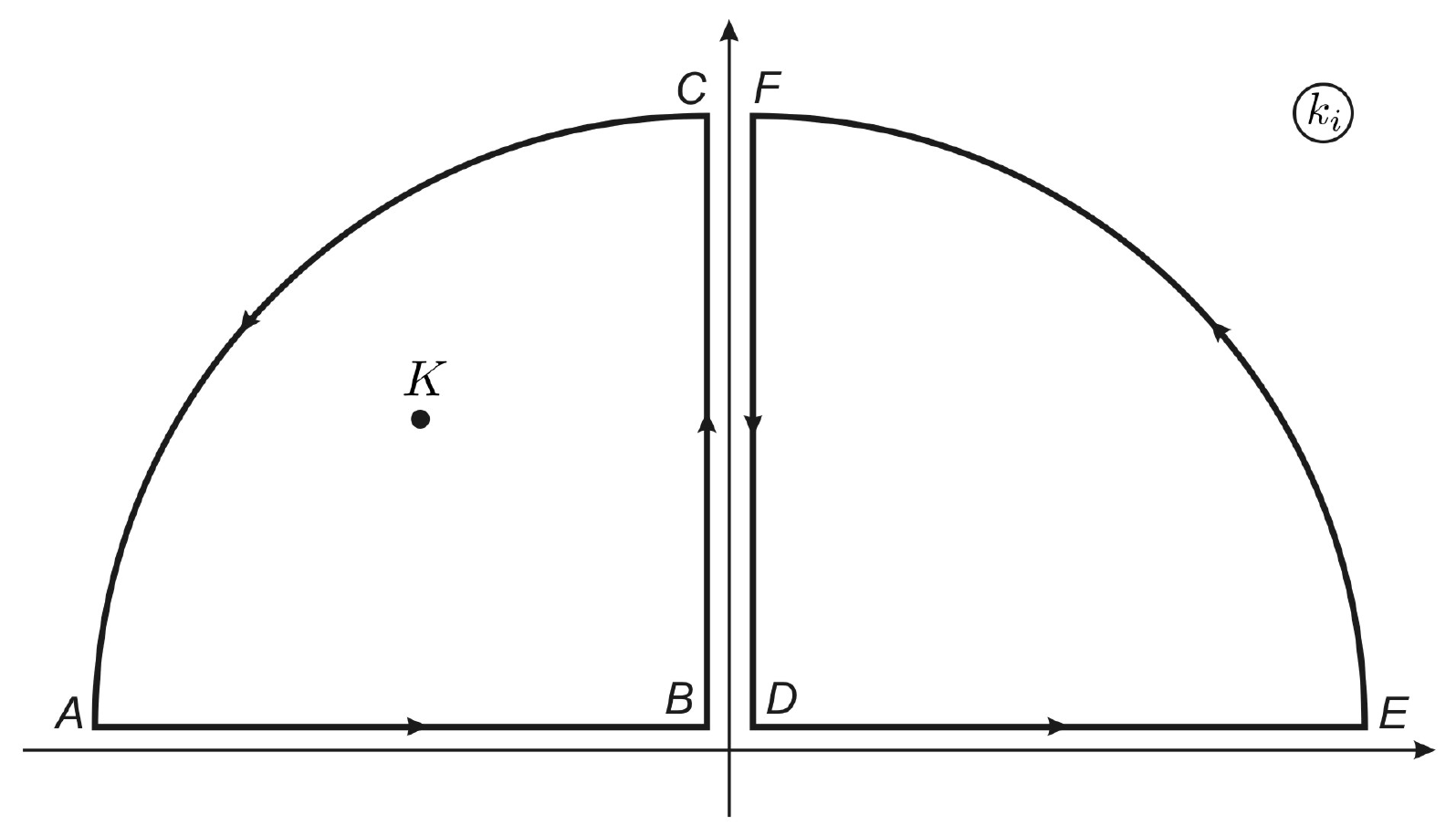}
\caption{To calculation of integrals standing in equations (\ref{14}) and
(\ref{14ab})}\label{int.pdf}
\end{figure}

To find the integrals needed let us specify a contribution that gives the pole
located on the upper half-plane of the complex number $k$. With this aim, we
divide this half-plane into two parts related to the positive and negative
values of the real part of $k$. As shows Figure \ref{int.pdf}, integrals
in Eq.(\ref{10}) can be rewritten as follows:
\begin{equation} \label{14}
 \begin{split}
\int\limits_{-\infty}^{+\infty}\frac{f(k)}{k-K}{\rm
d}k\equiv\int\limits_{AB}\frac{f(k)}{k-K}{\rm
d}k+\int\limits_{DE}\frac{f(k)}{k-K}{\rm
d}k\\=\oint\limits_{ABC}\frac{f(k)}{k-K}{\rm
d}k-\left[\int\limits_{BC}\frac{f(k)}{k-K}{\rm
d}k+\int\limits_{CA}\frac{f(k)}{k-K}{\rm
d}k\right]\\+\oint\limits_{DEF}\frac{f(k)}{k-K}{\rm
d}k-\left[\int\limits_{EF}\frac{f(k)}{k-K}{\rm
d}k+\int\limits_{FD}\frac{f(k)}{k-K}{\rm d}k\right]\\
=\oint\limits_{ABC}\frac{f(k)}{k-K}{\rm
d}k+\oint\limits_{DEF}\frac{f(k)}{k-K}{\rm d}k\\
-\left[\int\limits_{BC}\frac{f(k)}{k-K}{\rm
d}k+\int\limits_{FD}\frac{f(k)}{k-K}{\rm
d}k\right]-\left[\int\limits_{CA}\frac{f(k)}{k-K}{\rm
d}k+\int\limits_{EF}\frac{f(k)}{k-K}{\rm d}k\right].
 \end{split}
\end{equation}
With tending radiuses of the arcs $CA$ and $EF$ to infinity, both integrals in
the last square brackets disappear. On the other hand, when both half-axes $BC$
and $FD$ tend one to another, one has $\int_{BC}=-\int_{FD}$, so that terms in
the square brackets standing before are cancelled also. Moreover, the integral
over the contour $DEF$ equals zero because this contour does not envelop any
pole. As a result, we obtain
\begin{equation} \label{14ab}
\int\limits_{-\infty}^{+\infty}\frac{f(k)}{k-K}{\rm
d}k=\oint\limits_{ABC}\frac{f(k)}{k-K}{\rm d}k={\rm sgn}(\Im K)2\pi{\rm i}f(K)
\end{equation}
where the last equality is due to the residue theorem.

Finally, making use of the Cauchy integral (\ref{14ab}) yields the probability
distribution (\ref{10}) in the form
\begin{equation} \label{16}
P\left(x_1,x_2\right)=F_1^{\frac{2-\alpha}{\alpha-1}}P_1{\rm e}^{-{\rm
i}\left(K_1x_1-\phi_1\right)}+F_2^{\frac{2-\alpha}{\alpha-1}}P_2{\rm e}^{-{\rm
i}\left(K_2x_2-\phi_2\right)}
\end{equation}
where one denotes
\begin{equation} \label{17}
\begin{split}
P_{1,2}&\equiv\frac{J_{2,1}\left(K_{1,2}\right)}{(\alpha-1)|g_{2,1}|^{\frac{\alpha}{2(\alpha-1)}}
|m_{1,2}g_{1,2}|^{\frac{\alpha(3-\alpha)}{2(\alpha-1)}}},\\
\phi_{1,2}&\equiv\frac{3-\alpha}{\alpha-1}\varphi_{1,2}\left(\frac{\alpha}{2}\right)
+\frac{\pi}{2}\frac{2-\alpha}{\alpha-1}\left[{\rm sgn}\left(\Im
K_{1,2}\right)+2\theta\left(-\Re K_{1,2}\right)\right].
\end{split}
\end{equation}

\section{Discussion}\label{Sec.3}

Analytical consideration developed in previous Section allowed us to obtain the
probability distribution function (\ref{16}) that describes behaviour of
nonequilibrium steady-state stochastic system driven by the L\'evy
multiplicative noise with two degrees of freedom. Recently, we have studied
conditions of the limit cycle creation in stochastic Lorenz-type systems driven
by Gaussian noises \cite{Scripta}. Noise induced resonance was found
analytically to appear in non-equilibrium steady state if the fastest
variations displays a principle variable which is coupled with two different
degrees of freedom or more. The condition of this resonance appearance is
expressed formally in divergence of the probability distribution function being
inverse proportional to an effective force type of (\ref{9aa}) -- when this
force vanishes on a closed curve of phase plane, the system evolves along this
cycle with diverging probability density.

In opposite to such a dependence, the distribution function (\ref{16}) contains
the effective force (\ref{9aa}) in positive power $(2-\alpha)/(\alpha-1)$ only.
To this end, we can conclude the L\'evy flights suppress always a
quasi-periodical motion related to the limit cycle. That is main result of our
consideration. The corner stone of the difference between stochastic systems
driven by the L\'evy and Gaussian noises is that the L\'evy variation $\Delta
L\sim(\Delta t)^{1/\alpha}$ with the exponent $\alpha<2$ is negligible in
comparison with the Gaussian variation $\Delta W\sim(\Delta t)^{1/2}$ in the
$\Delta t\to 0$ limit.

It is interesting to note that above difference removes the problem of the
calculus choice \cite{2,23}. This problem is known to be caused by irregularity
of the time dependence $X(t)$ of stochastic variable (for the sake of
simplicity, we consider one-dimensional case again). Hence, in the integral of
equation of motion (\ref{1})
\begin{equation}
X(t)=\int\limits_0^t f\big(x(t')\big){\rm d}t'+\int\limits_{L(0)}^{L(t)}
g\big(x(\tilde{t'})\big){\rm d}L(t')
 \label{18}
\end{equation}
we should take the noise amplitude $g\big(x(\tilde{t'})\big)$ at the time
moment
\begin{equation}
\tilde{t'}=t'+\lambda\Delta t';\quad\lambda\in[0,1],\ \Delta t'\to 0
 \label{19}
\end{equation}
that does not coincide with the integration time $t'$ due to a parameter
$\lambda\in[0,1]$ whose value fixes calculus choice (for example, the magnitude
$\lambda=1/2$ relates to the Stratonovich case) \cite{2,23}. With accounting
equations (\ref{19}) and (\ref{1}), one obtains
\begin{equation} \label{20}
\begin{split}
g\big(x(\tilde{t})\big)&\simeq g\big(x(t)\big)+\lambda g'\big(x(t)\big)\Delta
X(t)\\ &\simeq g\big(x(t)\big)+\lambda g'\big(x(t)\big)f\big(x(t)\big)\Delta
t+\lambda g'\big(x(t)\big)g\big(x(t)\big)\Delta L(t)
\end{split}
\end{equation}
where primes denote differentiation over argument $x$. Being inserted into
Eq.(\ref{18}), the first term in the last line of Eq.(\ref{20}) relates to
usual case of the \^{I}to calculus. Corresponding insertion of the second term
gives an addition whose order $\Delta L\cdot\Delta t\sim(\Delta
t)^{1+(1/\alpha)}\ll\Delta t$ is higher than one for the previous term (such a
situation is inherent in the Gaussian case as well). Finally, after insertion
of the last term of Eq.(\ref{20}) the last integrand in Eq.(\ref{18}) obtains
an addition of order $(\Delta L)^2\sim(\Delta t)^{2/\alpha}$. In special case
of the Gaussian noise $(\alpha=2)$, the order $2/\alpha$ of this addition
coincides with the same in the first integrand of Eq.(\ref{18}), that is
resulted in addition $\lambda g(x)g'(x)$ to the physical force $f(x)$.
Principally different situation is realized for the L\'evy stable process, when
the index $\alpha<2$ and above addition should be suppressed in comparison with
physical force because $(\Delta t)^{2/\alpha}\ll\Delta t$.

\section*{Appendix A. Derivation of Fokker-Planck equation
for the L\'evy \break multiplicative noises}\label{Sec.4}
 \def\theequation{{A}.\arabic{equation}}
 \setcounter{equation}{0}

Following to the line of Ref. \cite{Yanovsky}, we start with
consideration of one-dimensional L\'evy process $X(t)$ whose
Chapman-Kolmogorov equation
\begin{equation}
p(x,t+{\rm d}t|x_0,t_0)=\int{\rm d}y~ p(x,t+{\rm
d}t|y,t)p(y,t|x_0,t_0)
 \label{1a}
\end{equation}
connects transition probabilities taken in intermediate positions
$y$ related to time $t$. According to the definition
\begin{equation}
p(k,t+{\rm d}t|y,t):={\rm e}^{{\rm d}K_X(k,{\rm d}t|y,t)},
 \label{3a}
\end{equation}
the inverse Fourier transform
\begin{equation}
p(x,t+{\rm d}t|y,t)=\int\frac{{\rm d}k}{2\pi}{\rm e}^{-{\rm
i}k(x-y)}p(k,t+{\rm d}t|y,t)
 \label{2a}
\end{equation}
is expressed in terms of the elementary cumulant ${\rm
d}K_X(k,{\rm d}t|y,t)$ of the characteristic function of the
stochastic process $X(t)$. Then, with using the ${\rm d}t\to 0$
limit and the identity
\begin{equation}
p(x,t|x_0,t_0)=\int{\rm d}y~ p(y,t|x_0,t_0)\int\frac{{\rm
d}k}{2\pi}{\rm e}^{-{\rm i}k(x-y)},
 \label{4a}
\end{equation}
the equation (\ref{1a}) arrives at the chain of equalities:
\begin{equation}
\begin{split}
p(x,t+{\rm d}t|x_0,t_0)-p(x,t|x_0,t_0)\\ =\int{\rm d}y~
p(y,t|x_0,t_0)\int\frac{{\rm d}k}{2\pi}{\rm e}^{-{\rm
i}k(x-y)}[{\rm e}^{{\rm d}K_X(k,{\rm d}t|y,t)}-1]\\ \simeq\int{\rm
d}y~ p(y,t|x_0,t_0)\int\frac{{\rm d}k}{2\pi}{\rm e}^{-{\rm
i}k(x-y)}{\rm d}K_X(k,{\rm d}t|y,t)\\=\int{\rm d}y~ {\rm
d}K_X(x-y,t)p(y,t|x_0,t_0)\equiv{\rm d}K_X(x,t)\star
p(x,t|x_0,t_0).
 \label{5a}
\end{split}
\end{equation}
Here, $\star$ denotes the convolution of the inverse Fourier
transform
\begin{equation}
{\rm d}K_X(x-y,t)=\int\frac{{\rm d}k}{2\pi}{\rm d}K_X(k,{\rm
d}t|y,t){\rm e}^{-{\rm i}k(x-y)}.
 \label{6a}
\end{equation}
As a result, with accounting the definition
\begin{equation}
{\rm d}K_X(x,t):=\mathcal{L}(x){\rm d}t,
 \label{7a}
\end{equation}
the equalities (\ref{5a}) yield the Fokker-Planck equation
\begin{equation}
\frac{\partial}{\partial t}p(x,t|x_0,t_0)=\mathcal{L}(x)\star
p(x,t|x_0,t_0).
 \label{8a}
\end{equation}

To obtain the explicit form of the increment $\mathcal{L}(x)$ let
us consider initially the L\'evy process $L(t)$ itself. The
elementary characteristic function related
\begin{equation}
\left<{\rm e}^{{\rm i}k{\rm d}L}\right>:={\rm e}^{{\rm
d}K_L(k,{\rm d}t|y,t)} \label{9a}
\end{equation}
is determined by the cumulant
\begin{equation}
{\rm d}K_L(k,{\rm d}t|y,t):=\Lambda(k){\rm d}t
 \label{10a}
\end{equation}
with the L\'evy increment \cite{Levy}
\begin{equation}
\Lambda(k)={\rm i}k\gamma-D|mk|^\alpha{\rm e}^{-{\rm
i}\varphi(\alpha)}
 \label{11a}
\end{equation}
where asymmetry angle $\varphi$ and modulus $m$ are determined by
Eqs. (\ref{12}). The elementary characteristic function of the
principle process ${\rm d}X=f{\rm d}t+g{\rm d}L$ is written as
follows:
\begin{equation}
{\rm e}^{{\rm d}K_X(k,{\rm d}t|y,t)}:=\left<{\rm e}^{{\rm i}k{\rm
d}X}\right>={\rm e}^{{\rm i}kf{\rm d}t}\left<{\rm e}^{{\rm i}(kg){\rm
d}L}\right>={\rm e}^{{\rm i}kf{\rm d}t}{\rm e}^{{\rm d}K_L(gk,{\rm d}t|y,t)}
 \label{12a}
\end{equation}
where Eq.(\ref{9a}) is taken into account. Similarly to the
definition (\ref{10a}), the elementary cumulant
\begin{equation}
{\rm d}K_X(k,{\rm d}t|y,t):=\mathcal{L}(k,x){\rm d}t
 \label{13a}
\end{equation}
is determined by the increment
\begin{equation}
\mathcal{L}(k,x)={\rm i}kf(x)+\Lambda\big(g(x)k\big)
 \label{14a}
\end{equation}
whose explicit form reads \cite{Yanovsky,Denisov}
\begin{equation}
\mathcal{L}(k,x)={\rm i}k\left[f(x)+\gamma
g(x)\right]-|mg(x)k|^\alpha{\rm e}^{-{\rm i}\varphi(\alpha)}.
 \label{15a}
\end{equation}
Hereafter, we renormalize the noise amplitude $g(x)$ to suppress
the scale factor $D$.

The probability distribution function
\begin{equation}
P(x,t)=\int{\rm d}x_0~p(x,t|x_0,t_0)P(x_0,t_0)
 \label{16a}
\end{equation}
is determined by the Fokker-Planck equation
\begin{equation}
\frac{\partial}{\partial t}P(x,t)=\mathcal{L}(x)\star
P(x,t)\equiv\int{\rm d}y~\mathcal{L}(x-y,x)P(y,t)
 \label{17a}
\end{equation}
following from Eq.(\ref{8a}). After using the Fourier transform
\begin{equation}
\widetilde{P}(k,t)\equiv\mathcal{F}\{P(y,t)\}(k,t)=\int{\rm
d}y~P(y,t){\rm e}^{{\rm i}ky}
 \label{18a}
\end{equation}
this equation takes the convenient form
\begin{equation}
\frac{\partial}{\partial
t}\widetilde{P}(k,t)=\mathcal{L}(k,x)\widetilde{P}(k,t)
 \label{19a}
\end{equation}
with the kernel (\ref{15a}). It is worthwhile to note this kernel,
being the Fourier transform inverse to Eq.(\ref{6a}) with respect
to the coordinate difference $x-y$, depends on the coordinate $x$
through both force $f(x)$ and multiplicative noise amplitude
$g(x)$.

With accounting the relation
\begin{equation}
\frac{\partial^\alpha}{\partial|x|^\alpha}h(x)=
-\mathcal{F}^{-1}\left\{|k|^\alpha\tilde{h}(k)\right\}
 \label{20a}
\end{equation}
for the Riesz derivative with respect to an arbitrary function
$h(x)$, Eqs. (\ref{19a}) and (\ref{15a}) arrive at the following
form of the fractional Fokker-Planck equation
\cite{Yanovsky,Denisov}
\begin{equation} \label{21a}
\begin{split}
\frac{\partial}{\partial t}P(x,t)=& -\frac{\partial}{\partial
x}\left[f(x)+\gamma g(x)\right]P(x,t)\\
&+\left[\frac{\partial^\alpha}{\partial|x|^\alpha}
+\beta\tan\left(\frac{\pi\alpha}{2}\right)\frac{\partial}{\partial
x}
\frac{\partial^{\alpha-1}}{\partial|x|^{\alpha-1}}\right]|g(x)|^\alpha
P(x,t).
\end{split}
\end{equation}
In symbolic form, many-dimensional generalization of this equation
for a symmetric L\'evy flight reads:
\begin{equation} \label{22a}
\frac{\partial}{\partial
t}P(\vec{x},t)=-\nabla\left[\vec{f}(\vec{x})+
\hat{g}(\vec{x})\cdot\vec{\gamma}\right]P(\vec{x},t)
-\left[-\hat{\Delta}:\vec{g}(\vec{x})\vec{g}(\vec{x})\right]^{\alpha/2}P(\vec{x},t).
\end{equation}
Here, every dot denotes the summation over indexes $i=1,2$ and the axes $x_1$,
$x_2$ forming pseudovector $\vec{x}$ are chosen so that the noise amplitude
matrix $\hat{g}$ takes the diagonal form $g_{ij}=g_i\delta_{ij}$ whose elements
form the pseudovector $\vec{g}$. In the component form, one has the continuity
equation
\begin{equation} \label{23a}
\frac{\partial}{\partial
t}P(\vec{x},t)+\sum\limits_i\frac{\partial}{\partial
x_i}J_i(\vec{x})=0
\end{equation}
with the probability flux
\begin{equation} \label{24a}
J_i(\vec{x})=\left\{\left[f_i(\vec{x})+g_i(\vec{x})\gamma_i\right]
+\frac{\partial^{\frac{\alpha}{2}-1}}{\partial
x_i^{\frac{\alpha}{2}-1}}\sum\limits_j\left(-\frac{\partial}{\partial
x_j}\right)^{\frac{\alpha}{2}}\left[g_i(\vec{x})g_j(\vec{x})\right]^{\frac{\alpha}{2}}\right\}
P(\vec{x}).
\end{equation}
In generalized case of non-symmetric L\'evy flights, the Fourier transforms of
the flux components are written in the explicit form
\begin{equation} \label{25a}
\begin{split}
J_1&=\left\{\left(f_1+g_1\gamma_1\right)+{\rm
i}|m_1g_1|^{\frac{\alpha}{2}}{\rm e}^{-{\rm
i}\varphi_1\left(\frac{\alpha}{2}\right)}|k_1|^{\frac{\alpha}{2}-2}k_1\right.\\
&\left.\times \left[\left|m_1g_1k_1\right|^{\frac{\alpha}{2}}{\rm
e}^{-{\rm
i}\varphi_1\left(\frac{\alpha}{2}\right)}+\left|m_2g_2k_2\right|^{\frac{\alpha}{2}}{\rm
e}^{-{\rm
i}\varphi_2\left(\frac{\alpha}{2}\right)}\right]\right\}\widetilde{P},\\
J_2&=\left\{\left(f_2+g_2\gamma_2\right)+{\rm
i}|m_2g_2|^{\frac{\alpha}{2}}{\rm e}^{-{\rm
i}\varphi_2\left(\frac{\alpha}{2}\right)}|k_2|^{\frac{\alpha}{2}-2}k_2\right.\\
&\left.\times\left[\left|m_1g_1k_1\right|^{\frac{\alpha}{2}}{\rm
e}^{-{\rm i}\varphi_1\left(\frac{\alpha}{2}\right)}+
\left|m_2g_2k_2\right|^{\frac{\alpha}{2}}{\rm e}^{-{\rm
i}\varphi_2\left(\frac{\alpha}{2}\right)}\right]\right\}\widetilde{P}
\end{split}
\end{equation}
where we use the asymmetry parameters (\ref{12}).

\section*{Appendix B. Consideration of the L\'evy processes
within direct \break stochastic space}\label{Sec.5}
 \def\theequation{{B}.\arabic{equation}}
 \setcounter{equation}{0}

After inverse Fourier transformation, the components (\ref{7}) and
(\ref{8}) of the stationary probability flux are written as
follows:
\begin{equation} \nonumber
\begin{split}
\left\{\left(f_1+g_1\gamma_1\right)+\frac{\partial^{\frac{\alpha}{2}-1}}{\partial
x_1^{\frac{\alpha}{2}-1}}\left[\left(-\frac{\partial}{\partial
x_1}\right)^{\frac{\alpha}{2}}g_1^\alpha+\left(-\frac{\partial}{\partial
x_2}\right)^{\frac{\alpha}{2}}\left(g_1g_2\right)^{\frac{\alpha}{2}}\right]\right\}P
=J_1^{(0)}(x_2),\\
\left\{\left(f_2+g_2\gamma_2\right)+\frac{\partial^{\frac{\alpha}{2}-1}}{\partial
x_2^{\frac{\alpha}{2}-1}}\left[\left(-\frac{\partial}{\partial
x_1}\right)^{\frac{\alpha}{2}}\left(g_2g_1\right)^{\frac{\alpha}{2}}
+\left(-\frac{\partial}{\partial
x_2}\right)^{\frac{\alpha}{2}}g_2^\alpha\right]\right\}P
=J_2^{(0)}(x_1).
\end{split}
\end{equation}
Acting by the
$g_2^{\frac{\alpha}{2}}\frac{\partial^{\frac{\alpha}{2}-1}}{\partial
x_2^{\frac{\alpha}{2}-1}}$ operator on the first of these
equations and the
$g_1^{\frac{\alpha}{2}}\frac{\partial^{\frac{\alpha}{2}-1}}{\partial
x_1^{\frac{\alpha}{2}-1}}$ operator -- on the second, one obtains
\begin{equation} \label{1b}
\begin{split}
g_2^{\frac{\alpha}{2}}\frac{\partial^{\frac{\alpha}{2}-1}}{\partial
x_2^{\frac{\alpha}{2}-1}}\frac{\partial^{\frac{\alpha}{2}-1}}{\partial
x_1^{\frac{\alpha}{2}-1}}\left[\left(-\frac{\partial}{\partial
x_1}\right)^{\frac{\alpha}{2}}g_1^\alpha+\left(-\frac{\partial}{\partial
x_2}\right)^{\frac{\alpha}{2}}(g_1g_2)^{\frac{\alpha}{2}}\right]
P\\=g_2^{\frac{\alpha}{2}}\frac{\partial^{\frac{\alpha}{2}-1}}{\partial
x_2^{\frac{\alpha}{2}-1}}
\left[J_1^{(0)}-\left(f_1+g_1\gamma_1\right)P\right],\\
g_1^{\frac{\alpha}{2}}\frac{\partial^{\frac{\alpha}{2}-1}}{\partial
x_1^{\frac{\alpha}{2}-1}}\frac{\partial^{\frac{\alpha}{2}-1}}{\partial
x_2^{\frac{\alpha}{2}-1}}\left[\left(-\frac{\partial}{\partial
x_1}\right)^{\frac{\alpha}{2}}\left(g_2g_1\right)^{\frac{\alpha}{2}}
+\left(-\frac{\partial}{\partial
x_2}\right)^{\frac{\alpha}{2}}g_2^\alpha \right]
P\\=g_1^{\frac{\alpha}{2}}\frac{\partial^{\frac{\alpha}{2}-1}}{\partial
x_1^{\frac{\alpha}{2}-1}}
\left[J_2^{(0)}-\left(f_2+g_2\gamma_2\right)P\right].
\end{split}
\end{equation}
Subtracting above equalities term by term, one arrives at the
fractional differential equation
\begin{equation} \label{2b}
\begin{split}
\left[\left(f_1+g_1\gamma_1\right)g_2^{\frac{\alpha}{2}}\frac{\partial^{\frac{\alpha}{2}-1}}
{\partial
x_2^{\frac{\alpha}{2}-1}}-\left(f_2+g_2\gamma_2\right)g_1^{\frac{\alpha}{2}}
\frac{\partial^{\frac{\alpha}{2}-1}} {\partial
x_1^{\frac{\alpha}{2}-1}}\right]P+G\left(x_1,x_2\right)P\\
=g_2^{\frac{\alpha}{2}}\frac{\partial^{\frac{\alpha}{2}-1}}{\partial
x_2^{\frac{\alpha}{2}-1}}J_1^{(0)}\left(x_2\right)-g_1^{\frac{\alpha}{2}}\frac{\partial^{\frac{\alpha}{2}-1}}
{\partial x_1^{\frac{\alpha}{2}-1}}J_2^{(0)}\left(x_1\right)
\end{split}
\end{equation}
where one denotes the function
\begin{equation} \label{3b}
\begin{split}
G\left(x_1,x_2\right)\equiv
g_2^{\frac{\alpha}{2}}\frac{\partial^{\frac{\alpha}{2}-1}}{\partial
x_2^{\frac{\alpha}{2}-1}}\frac{\partial^{\frac{\alpha}{2}-1}}{\partial
x_1^{\frac{\alpha}{2}-1}}\left[\left(-\frac{\partial}{\partial
x_1}\right)^{\frac{\alpha}{2}}g_1^\alpha+\left(-\frac{\partial}{\partial
x_2}\right)^{\frac{\alpha}{2}}(g_1g_2)^{\frac{\alpha}{2}}\right]\\
-g_1^{\frac{\alpha}{2}}\frac{\partial^{\frac{\alpha}{2}-1}}{\partial
x_1^{\frac{\alpha}{2}-1}}\frac{\partial^{\frac{\alpha}{2}-1}}{\partial
x_2^{\frac{\alpha}{2}-1}}\left[\left(-\frac{\partial}{\partial
x_1}\right)^{\frac{\alpha}{2}}\left(g_2g_1\right)^{\frac{\alpha}{2}}
+\left(-\frac{\partial}{\partial
x_2}\right)^{\frac{\alpha}{2}}g_2^\alpha
\right]\\
+\left[g_2^{\frac{\alpha}{2}}\frac{\partial^{\frac{\alpha}{2}-1}}{\partial
x_2^{\frac{\alpha}{2}-1}}\left(f_1+g_1\gamma_1\right)-
g_1^{\frac{\alpha}{2}}\frac{\partial^{\frac{\alpha}{2}-1}}{\partial
x_1^{\frac{\alpha}{2}-1}}\left(f_2+g_2\gamma_2\right)\right].
\end{split}
\end{equation}
For the Gauss processes $(\alpha=2)$, the differential equation
(\ref{2b}) is reduced to the algebraic one to give the probability
distribution function that has been found in our previous work
\cite{Scripta}. However, in general case $\alpha\leq 2$, solution
of the fractional differential equation (\ref{2b}) arrives at a
complicated problem, so that we are obliged to use the Fourier
representation in Section \ref{Sec.2}.

It is worthwhile to note finally the consistency condition
(\ref{condition}) takes the form
\begin{equation} \label{4b}
\begin{split}
\left[\frac{\partial}{\partial
x_1}\left(f_1+g_1\gamma_1\right)-\left(-\frac{\partial}{\partial
x_1}\right)^{\alpha}|g_1|^\alpha\right]J_2^{(0)}(x_1)\\
=\left[\frac{\partial}{\partial
x_2}\left(f_2+g_2\gamma_2\right)-\left(-\frac{\partial}{\partial
x_2}\right)^{\alpha}|g_2|^\alpha\right]J_1^{(0)}(x_2)
\end{split}
\end{equation}
within the inverse Fourier representation where one takes $\varphi_i=0$ and
$m_i=1$, for the simplicity. The equation (\ref{4b}) connects explicitly the
probability flux components $J_{2,1}(x_{1,2})$, being arbitrary functions, with
given dependencies of both forces $f_1(x_1,x_2)$, $f_2(x_1,x_2)$ and
multiplicative amplitudes $g_1(x_1,x_2)$, $g_2(x_1,x_2)$, respectively.


\begin{thebibliography}{39}

\bibitem{2} H. Horsthemke, R. Lefever, Noise Induced Transitions,
Springer-Verlag, Berlin, 1984.

\bibitem{2a} P. Reimann, Brownian motors: noisy transport far from
equilibrium, Phys. Rep. 361 (2002) 57–-265.

\bibitem{3a} C. Van den Broeck, J.M.R. Parrondo, R. Toral, Noise-Induced
Nonequilibrium Phase Transition, Phys. Rev. Lett. 73 (1994)
3395--3398; Mean field model for spatially extended systems in the
presence of multiplicative noise, Phys. Rev. E 49 (1994)
2639--2643.

\bibitem{3} A.I. Olemskoi, D.O. Kharchenko, I.A. Knyaz', Phase transitions
induced by noise cross-correlations, Phys. Rev. E 71 (2005)
041101.

\bibitem{4a} R. Benzi, A. Sutera, A. Vulpiani, The mechanism of stochastic resonance, J. Phys.
A: Math. Gen. 14 (1981) L453--L457.

\bibitem{4} L. Gammaitoni, P. H\"anggi, P. Jung, F. Marchesoni, Stochastic resonance, Rev. Mod. Phys. 70 (1998)
223--287.

\bibitem{a} J. Buceta, M. Ibaes, J.M. Sancho, Katja Lindenberg, Noise-driven mechanism
for pattern formation, Phys. Rev. E 67 (2003) 021113.

\bibitem{b} M.C. Cross, P.C. Hohenberg, Pattern formation
outside of equilibrium, Rev. Mod. Phys. 65 (1993) 851--1112.

\bibitem{c} F. Julicher, A. Ajdari, J. Prost, Modeling molecular motors, Rev. Mod. Phys. 69 (1997) 1269--1282.

\bibitem{SSG} F. Sagu\'es, J.M. Sancho, J. Garcia-Ojalv, Spatiotemporal order
out of noise, Rev. Mod. Phys. 79 (2007) 829--883.

\bibitem{e} R.L. Kautz, Chaos, thermal noise in the rf-biased Josephson junction, J. Appl. Phys. 58 (1985)
424--440.

\bibitem{f} F. Arecchi, R. Badii, A. Politi, Generalized multistability and
noise-induced jumps in a nonlinear dynamical system, Phys. Rev. A 32 (1985) 402--408.

\bibitem{d} J.B. Gao, Wen-wen Tung, Nageswara Rao, Noise-Induced
Hopf-Bifurcation-Type Sequence and Transition to Chaos in the Lorenz
Equations, Phys. Rev. Lett. 89 (2002) 254101.

\bibitem{g} Omar Osenda, Carlos B. Briozzo, Manuel O. Caceres, Stochastic
Lorenz model for periodically driven Rayleigh-Be'nard convection, Phys.
Rev. E. 55 (1997) R3824--R3827.

\bibitem{11} D. Alonso, A.J. McKane, M. Pascual, Stochastic amplification in epidemics, J. R. Soc. Interface 4 (2007) 575--582.

\bibitem{12} M. Simoes, M.M. Telo da Gama, A. Nunes, Stochastic fluctuations in epidemics on networks, J. R. Soc. Interface 5 (2008)
555--566.

\bibitem{13} R. Kuske, L.F. Gordillo, P. Greenwood, Sustained oscillations via
coherence resonance in SIR, J. Theor. Biol. 245 (2007)
459--469.

\bibitem{5} A.J. McKane, T.J. Newman, Predator-Prey Cycles
from Resonant Amplification of Demographic Stochasticity, Phys. Rev.
Lett. 94 (2005) 218102.

\bibitem{6} M. Pineda-Krch, H.J. Blok, U. Dieckmann, M. Doebeli, A tale of two cycles – Distinguishing between true limit
cycles and quasi-cycles in finite predator-prey populations, Oikos 116 (2007) 53--64.

\bibitem{10} M.S. de la Lama, I.G. Szendro, J.R. Iglesias, H.S. Wio, Van Kampen's expansion approach in an opinion formation model, Eur. Phys. J. B 51 (2006)
435--442.

\bibitem{15} D. Gonze, J. Halloy, P. Gaspard, Biochemical clocks and
molecular noise: Theoretical study of robustness factors, J. Chem. Phys. 116 (2002) 10997--11010.

\bibitem{16} A.J. McKane, J.D. Nagy, T.J. Newman, M.O. Stefanini, Amplified Biochemical Oscillations in Cellular Systems, J. Stat. Phys. 128 (2007)
165--191.

\bibitem{14} M. Scott, B. Ingalls, M. Kaern, Estimations of
intrinsic and extrinsic noise in models of nonlinear genetic networks, Chaos 16 (2006) 026107.

\bibitem{9} E. Ben-Naim, P.L. Krapivsky, Finite-size fluctuations
in interacting particle systems, Phys. Rev. E 69 (2004) 046113.

\bibitem{17} B.D. Hassard, N.D. Kazarinoff, Y.-H. Wan, Theory and Applications of
Hopf Bifurcation, Cambridge University Press, Cambridge, 1981.

\bibitem{18} R. Grimshaw, Nonlinear Ordinary Differential Equations, Blackwell, Oxford, 1990.

\bibitem{19} M.S. Bartlett, Measles periodicity and community size, J. R. Stat. Soc. A
120 (1957) 48--70.

\bibitem{20} A.J. McKane, T.J. Newman, Predator-Prey Cycles
from Resonant Amplification of Demographic Stochasticity, Phys. Rev.
Let. 94 (2005) 218102.

\bibitem{21a} A. Pikovsky, J. Kurths, Coherence Resonance in a
Noise-Driven Excitable System, Phys. Rev. Let. 78 (1997)
775--778.

\bibitem{21b} B. Lindner, L. Schimansky-Geier, Coherence and stochastic
resonance in a two-state system, Phys. Rev. E 61 (2000)
6103--6110.

\bibitem{21c} A. Neiman, P.I. Saparin, L. Stone, Coherence resonance at noisy
precursors of bifurcations in nonlinear dynamical systems, Phys. Rev. E
56 (1997) 270--273.

\bibitem{21d} H. Gang, T. Ditzinger, C.Z. Ning, H. Haken, Stochastic resonance without
external periodic force, Phys. Rev. Let. 71 (1993) 807--810.

\bibitem{rev} A.A. Dubkov, B. Spagnolo, V.V. Uchaikin, L\'evy Flight Superdiffusion: An Introduction,
Intern. Journ. of Bifurcation and Chaos 18 No. 9 (2008)
2649--2672.

\bibitem{0} A. Ichiki, M. Shiino, Phase transitions driven by L\'evy stable noise: exact solutions and
stability analysis of nonlinear fractional Fokker-Planck equations,
arXiv:cond-mat.stat-mech/0907.3782v1.

\bibitem{Yanovsky} D. Schertzer, M. Larchev$\hat{\rm e}$que, J. Duan, V.V. Yanovsky, S. Lovejoy, Fractional Fokker-Planck equation for nonlinear
stochastic differential equations driven by non-Gaussian L\'evy stable noise,
J. Math. Phys: Math. Gen. 41 (2001) 200--212.

\bibitem{Denisov} S.I. Denisov, W. Horsthemke, P. H\"anggi, Generalized
Fokker-Planck equation: Derivation and exact solutions, Eur. Phys. J. B
68 (2009) 567–-575.

\bibitem{23} H. Risken, The Fokker-Planck Equation, Springer-Verlag, Berlin, 1984.

\bibitem{Scripta} I.A. Shuda, S.S. Borysov, A.I. Olemskoi, Noise-induced oscillations in
non-equilibrium steady state systems, Physica Scripta 79 (2009)
065001.

\bibitem{Levy} P. L\'evy, Theorie de l'addition des variables Al\'eatoires, Gauthier-Villars, Paris, 1937.

\end{thebibliography}
\end{document}